\documentclass[aps,prb,reprint,showpacs,showkeys,amsmath]{revtex4-1}
\usepackage{hyperref}
\usepackage{graphicx}

\begin{document}

\title{Externally-driven transformations of vortex textures in flat submicrometer magnets}
\author{Andrzej Janutka}
\email{Andrzej.Janutka@pwr.wroc.pl}
\affiliation{Institute of Physics, Wroclaw University of Technology, Wybrze\.ze Wyspia\'nskiego 27, 50-370 Wroc{\l}aw, Poland}

\begin{abstract}

Two effects of oscillatory transformations of vortex textures in flat nanomagnets due to the application of 
 an external field or a spin-polarized electric current are analytically described with relevance to soft-magnetic  
 structures of submicrometer sizes (whose thickness is significantly bigger than the magnetostatic exchange length).
 These are changes of a domain wall (DW) structure in a long magnetic stripe (oscillations between a transverse DW,
 a vortex DW, and an antivortex DW) and periodic vortex-core reversals in a circular magnetic dot which are accompanied
 by oscillatory displacements of the vortex from the dot center. In nanostructures of smaller thicknesses (comparable
 to the exchange length), where nonlocal magnetostatic effects are very strong because of fast spatial variation
 of the magnetization, similar phenomena have been widely studied previously. Here, the dynamics is investigated
 within a local approach including magnetostatic field via boundary conditions on solutions
 to the Landau-Lifshitz-Gilbert equation only. Both the DWs in stripes and vortex states of the dot are treated
 as fragments of a cross-tie DW. Despite similarity of the cyclic transformations of the ordering to the dynamics
 of more strongly confined nanomagnets, details of motion (trajectories) of the vortices and antivortices (Bloch lines)
 of the textures under study are different, which is related to prohibition of rapid jumps of the polarization of Bloch lines.
 In addition to the magnetization rotation about the direction of magnetic field or current polarization,
 the evolution of textures is shown to relate to oscillatory changes of the direction of a cross-tie DW with respect
 to any arbitrary axis in the magnet plane accompanied by oscillations of the DW width.
 \end{abstract}


\maketitle
\newpage

\section{Introduction}

Application of external (longitudinal) magnetic field or voltage to the DW-containing ferromagnetic nanowire drives translational
 motion of the DW. When the intensity of the field or electric current exceeds critical value of the Walker breakdown,
 the DW translation is accompanied by cyclic transformations of the DW structure \cite{sch74}. In soft-magnetic nanostripes,
 strong magnetostatic field stabilizes textures (DWs) whose topological charges are pinned to the platelet
 boundary and whose topology gets transitions with changing thickness and width
 of the nanostripe \cite{mcm97,nak05,kla08}. There, the field-induced DW transformations are transitions
 between the so-called transverse and vortex DWs (antivortex DWs) \cite{thi05,kun06,lee07,hay08a}.   
 
In circular magnetic dots, magnetostatic field can form a centered vortex of magnetization \cite{pok00,shi00}.
 In the nanocontact with a spin valve structure, the dot conducts in-plane-polarized electric current in the out-of-plane
 direction, which leads to the transition of vortex (with increase of the current intensity) to the state of precession
 about the center of the dot \cite{puf07,pri07,len09,slu11}. At low current intensity, the vortex sustains displacement moving
 out the dot \cite{ish06}.

Electrically-induced dynamical transformations of planar spin-textures play the key role in magnetic nano-oscillators 
 for applications to microwave generation and sensing \cite{kis03,tho06,sil07}. In terms of design of DW-based
 storage and logic devices \cite{par08,hay08,all05}, the DW-structure transformation due to longitudinal magnetic
 field or electric current is an undesired effect which limits the efficiency range of velocities of
 the DW propagation \cite{hrk11}.

Since the ordering in soft-magnetic nanostructures is governed by the exchange as well as by long-range dipolar
 interactions (permalloys are the most popular ferromagnetic materials for the above applications), description
 of dynamical phenomena is nonlocal and remains a challenge. At present, micromagnetic simulations are the main source
 of knowledge on details of the texture evolution in spatially confined magnets. Analytical studies of the dynamics
 of DWs in nanostripes, \cite{tre08,cla08}, and of vortex-states in circular dots, \cite{khv09}, 
 use modifications of a Thiele method that treats the magnetic vortices as moving rigid objects \cite{thi73}.
 The spin-structure transformations are related to the motion of a number of vortices and antivortices, thus,
 the dynamical (Thiele) equations of the (anti)vortex-core positions constitute complex systems. Moreover,
 they contain semi-empirical (gyrotropic and viscosity) coefficients to be estimated.

With relevance to thick-enough planar nanomagnets, in the present paper, I analytically study the field- and current-induced
 transformations of DWs in ferromagnetic stripes solving the Landau-Lifshitz-Gilbert (LLG) equation instead of using the Thiele
 equations. Within a relatively-simple local approach, the magnetostatic effects are included via boundary conditions only.
 Also, I apply the present method to the vortex-state transformations in a circular magnetic dot. I treat the vortex-state
 in such a quasi-2D system as well as the DW in the magnetic stripe as fragments of an infinite spin texture called
 the cross-tie DW \cite{mid63,uhl09}, (recently this approach has been applied to study DW interactions in Ref. \cite{jan12}).

In contrast to the most frequently investigated nanoelements of ultimately-small thicknesses, in (thicker) systems under
 considerations, the exchange interactions dominate over the magnetostatic effects while the later ones
 are included as a perturbation \cite{met05}. In very flat systems, the dipolar interactions dominate while
 the exchange is thought of as a perturbation \cite{che05,che08}. Similar approach is used by popular codes for micromagnetic
 simulations that apply the fast Fourier transform to calculate the dipolar (magnetostatic) field, which limits their
 validity \cite{mcm97,hin00}. The crossover between both the interaction regimes is relevant to thickness of the magnetic element
 close to the critical length $l_{c}$ of a "macrospin approximation" defined in Ref. \cite{ber08} and estimated to be from
 the range $4l_{ex}\div8l_{ex}$, where $l_{ex}$ denotes magnetostatic exchange length ($l_{ex}\approx 5$nm for permalloys). 
 In elements whose spatial sizes are smaller than $l_{c}$ the density of exchange energy is too high to allow an inhomogeneous
 ordering unless the anisotropy is nonuniform \cite{don04}. Decrease of one of the spatial sizes below $l_{c}$ in presence of 
 strong local anisotropy can result in discontinuity of spatial variation (differentials) of the magnetization.

Solving the LLG equation for the thick-film regime, differently oriented static cross-tie DWs are found to be analogs
 of the transverse and vortex DWs of very thin stripes. Upon externally-induced driving, the DW textures oscillatory transform
 between these two basic configurations, however, details of the transitions between transverse and vortex states are found
 to differ from those in the thin stripes in terms of trajectories of the DW-texture defects (vortices and antivortices).
 Unlike most of available descriptions of the vortex dynamics in nanodots which deal with defects in the in-plane-ordered
 (curling) state, here, evolution of a vortex texture distorted from the plane at the dot rim is analyzed. The trajectories
 of the current-induced vortex translation are examined as well as trajectories of antivortices which appear due to enforced 
 reversals of the defect core.
 
In Sec. II, stationary DW solutions to the LLG equation in the 2D stripe geometry are found. 
 Externally-driven dynamics of such structures is analyzed in Sec. III. Sec. IV is devoted to study of the structure
 and dynamics of vortex states in circular dots. Conclusions are collected in Sec. V. 
 
\section{Domain-wall states in ferromagnetic stripe}

Let us consider stationary DW solutions to the LLG equation in 2D
\begin{eqnarray}
-\frac{\partial{\bf m}}{\partial t}=\frac{J}{M}{\bf m}\times\left(\frac{\partial^{2}{\bf m}}{\partial x^{2}}
+\frac{\partial^{2}{\bf m}}{\partial z^{2}}\right)
+\gamma{\bf m}\times{\bf H}
\nonumber\\
+\frac{\beta_{1}}{M}({\bf m}\cdot\hat{i}){\bf m}\times\hat{i}
-\frac{\alpha}{M}{\bf m}\times\frac{\partial{\bf m}}{\partial t}.
\label{LLG}
\end{eqnarray}
Here, $M=|{\bf m}|$, $J$ denotes the exchange constant, $\beta_{1}$ determines strength of the easy-axis anisotropy, 
 ${\bf H}=(H_{x},0,0)$ represents the external (longitudinal) magnetic field, thus, $\gamma$ denotes the gyromagnetic factor.
 Although, in permalloy-like soft-magnetic materials, bulk anisotropy is negligible, for generality of considerations, I admit
 the presence of the easy-axis anisotropy (e.g. due to wire deposition on substrate \cite{alo05}), however, assuming it to be weak
 compared to the exchange, $\beta_{1}\ll J/w^{2}$, where $w$ denotes the stripe width. In soft-magnetic stripes, the shape
 anisotropy due to dipolar interactions (vanishing of magnetostatic charges at the magnet surfaces) aligns the magnetization
 of homogeneously ordered domains onto the direction of long axis of the stripe $\hat{i}\equiv(1,0,0)$, \cite{kla08,ved07}.
 Therefore, I study solutions that satisfy the condition ${\rm lim}_{|x|\to\infty}{\bf m}=\pm(M,0,0)$. 

Since (\ref{LLG}) must be solved with the constraint on the magnetization length, it is comfortable to consider
 equivalent to (\ref{LLG}) equations of the unconstrained dynamics. Introducing $m_{\pm}\equiv m_{y}\pm{\rm i}m_{z}$, one represents
 the magnetization components with a pair of complex functions $g(x,z,t)$, $f(x,z,t)$ (secondary dynamical variables);
\begin{eqnarray}
m_{+}=\frac{2M}{f^{*}/g+g^{*}/f},
\hspace*{2em}
m_{x}=M\frac{f^{*}/g-g^{*}/f}{f^{*}/g+g^{*}/f},
\label{transform}
\end{eqnarray}
thus ensuring that $|{\bf m}|=M$. Insertion of (\ref{transform}) into (\ref{LLG}) leads,
 following the Hirota method for solving nonlinear differential equations \cite{bog80,kos90},
 to the trilinear equations of motion
\begin{eqnarray}
-f{\rm i}D_{t}f^{*}\cdot g&=&f\left[\alpha D_{t}+J(D_{x}^{2}+D_{z}^{2})\right]f^{*}\cdot g 
+Jg^{*}(D_{x}^{2}
\nonumber\\
&&+D_{z}^{2})g\cdot g-\left(\gamma H_{x}+\beta_{1}\right)|f|^{2}g,
\nonumber\\
-g^{*}{\rm i}D_{t}f^{*}\cdot g&=&g^{*}\left[\alpha D_{t}-J(D_{x}^{2}+D_{z}^{2})\right]f^{*}\cdot g
-Jf(D_{x}^{2}
\nonumber\\
&&
+D_{z}^{2})f^{*}\cdot f^{*}+\left(-\gamma H_{x}+\beta_{1}\right)|g|^{2}f^{*},
\label{secondary-eq}
\end{eqnarray}
where $D_{t}$, $D_{x}$, $D_{z}$ denote Hirota operators of differentiation 
\begin{eqnarray}
D_{x}^{n}&&b(x,z,t)\cdot c(x,z,t)\equiv
\nonumber\\
&&(\partial/\partial x-\partial/\partial x^{'})^{n}b(x,z,t)c(x^{'},z^{'},t^{'})|_{x=x^{'},z=z^{'},t=t^{'}}.
\nonumber
\end{eqnarray}
For ${\bf H}=0$, stationary single-DW solutions to (\ref{secondary-eq}) are of the form
\begin{eqnarray}
f=1,\hspace*{2em}g=u{\rm e}^{kx+qz},
\label{solution_1}
\end{eqnarray}
where 
\begin{eqnarray} 
k^{2}+q^{2}=\frac{\beta_{1}}{J}
\label{condition_1}
\end{eqnarray}
and ${\rm Re}k\neq 0$. I denote $k\equiv k^{'}+{\rm i}k^{''}$, $q\equiv q^{'}+{\rm i}q^{''}$, where 
 $k^{'('')}$, $q^{'('')}$ take real values, hence, (\ref{condition_1}) is equivalent to 
\begin{eqnarray} 
k^{'2}+q^{'2}-k^{''2}-q^{''2}=\frac{\beta_{1}}{J},
\nonumber\\
k^{'}k^{''}+q^{'}q^{''}=0.
\label{condition_1_prime}
\end{eqnarray}
Assuming one of the DW edges to be centered at $x=0$, (then $u={\rm e}^{{\rm i}\phi}$), the relevant magnetization
 profile [the single-DW solution to (\ref{LLG})] is written explicitly with 
\begin{subequations} 
\begin{eqnarray} 
m_{+}(x,z)&=&M{\rm e}^{{\rm i}(\phi+k^{''}x+q^{''}z)}{\rm sech}[k^{'}x+q^{'}z],
\label{profile1a}\\
m_{x}(x,z)&=&-M{\rm tanh}[k^{'}x+q^{'}z].
\label{profile1b}
\end{eqnarray}
\end{subequations} 
Let $q^{''}\neq 0$ since, in the opposite case, the DW states are similar to DWs in 1D ferromagnet and cannot exist
 in absence of the bulk anisotropy while I consider soft-magnetic systems with very weak anisotropy admitting
 the case $\beta_{1}=0$, \cite{kos90,jan11}. Defining $\theta\equiv{\rm arctan}(q^{'}/k^{'})$, via (\ref{condition_1_prime}),
 one finds $k^{''}=-q^{''}\tan(\theta)$ and $k^{'2}-q^{''2}=\beta_{1}/\{J[1+\tan^{2}(\theta)]\}$. Also, I assume the magnetization
 orderings on both the stripe edges to be similar, thus, the phase factor of the RHS of (\ref{profile1a}) changes by $n\pi$ along
 the DW line $k^{'}x+q^{'}z=0$ between it ends, where $n=1,2,\ldots$. It leads to the condition $k^{''}(-wq^{'}/k^{'})+q^{''}w=n\pi$
 and, finally, to $q^{''}=n\pi/\{w[1+\tan^{2}(\theta)]\}$. 

\begin{figure}
\unitlength 1mm
\begin{center}
\begin{picture}(175,152)
\put(0,-5){\resizebox{80mm}{!}{\includegraphics{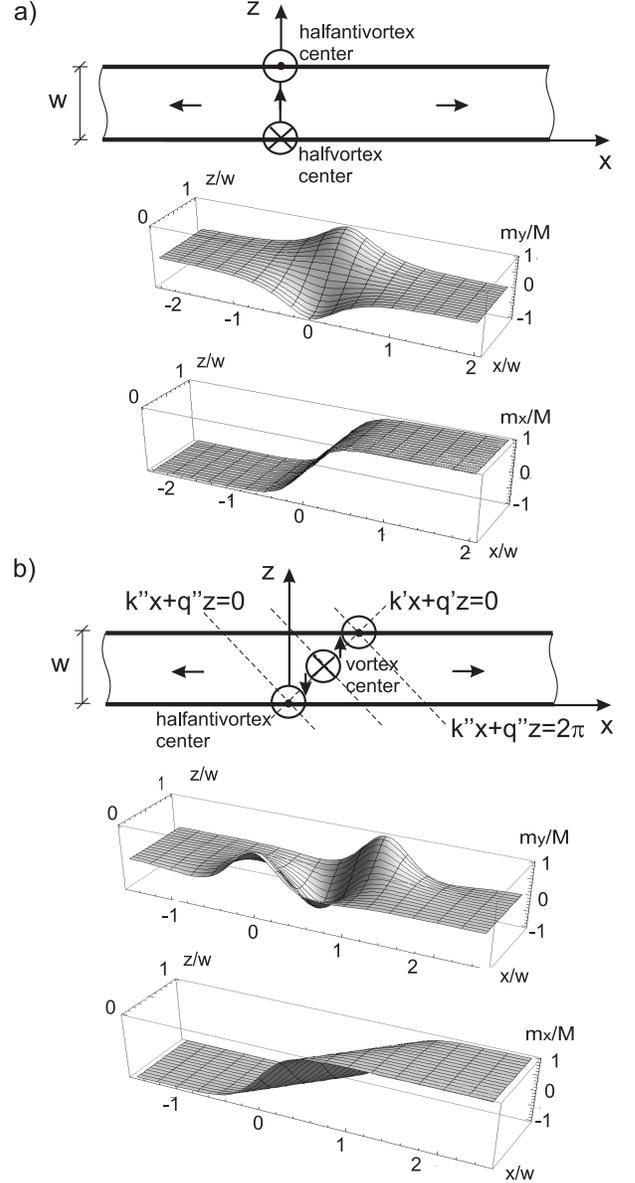}}}\end{picture}
\end{center}
\caption{DW configurations; a) a transverse DW, b) a vortex DW. In the upper draws, arrows indicate magnetization alignment.}
\end{figure}

Additional boundary condition is related to minimization of the surface (magnetostatic) energy and 
 it discriminates between different values of $\phi$, $n$, and $\theta$.
 I evaluate the energy of the DW with dependence on these parameters using 
 the Hamiltonian ${\cal H}={\cal H}_{0}+{\cal H}_{Z}$, where
\begin{eqnarray}
{\cal H}_{0}&=&\frac{J}{2M}\left(\bigg|\frac{\partial{\bf m}}{\partial x}\bigg|^{2}
+\bigg|\frac{\partial{\bf m}}{\partial z}\bigg|^{2}\right)
+\frac{\beta_{1}}{2M}\left[M^{2}-({\bf m}\cdot\hat{i})^{2}\right],
\nonumber\\
{\cal H}_{Z}&=&-\gamma{\bf H}\cdot{\bf m},
\label{hamiltonian}
\end{eqnarray}
(${\cal H}_{Z}$ denotes its Zeeman part). Total energy of the DW $E=E_{0}+E_{Z}+E_{B}$ is the sum 
 of the bulk energy $E_{0}+E_{Z}$, defined by 
 $E_{0(Z)}\equiv\int_{-\infty}^{\infty}\int_{0}^{w}{\cal H}_{0(Z)}{\rm d}z{\rm d}x$, 
 and of the boundary energy $E_{B}$ which is of the magnetostatic origin.

Since evaluating the magnetostatic energy of the stripe within phenomenological approach is a complex mathematical problem
 whose systematic solution is beyond the scope of this paper, I search for energy of the stripe boundary using a heuristic
 argumentation. The formula for $E_{B}$ is determined referring to a theorem by Carbou who proved that the magnetostatic
 energy of any ferromagnetic element of finite thickness $\tau$; $\sim1/\lambda^{2}\int_{S}({\bf m}\cdot{\bf n})^{2}{\rm d}s$
 tends to $1/\Lambda_{2}\int_{\partial S}({\bf m}\cdot{\bf n}^{'})^{2}{\rm d}l$ with $\tau\to 0$, \cite{car01,koh05}. Here, 
 $S$ denotes the surface of the bulk ferromagnet and $\partial S$ denotes the boundary of the base of its solid,
 ${\bf n}$ is normal to the magnet surface, ${\bf n}^{'}$ denotes the unitary vector outward
 to the line of the base boundary. The coefficient $\Lambda_{2}$ scales with $\tau$ and with a diameter $w$ following
 $\Lambda_{2}\sim\lambda^{2}/[\tau|\log(\tau/w)|]$, (with relevance to a stripe, $w$ represents
 its width \cite{che05,che08}). One has to notice that the Carbou theorem is not strictly applicable to systems
 with open boundaries, (infinite stripes), however, it shows some tendency in ordering at the stripe edges.
 In particular, it indicates that the magnetostatic interactions in flat magnets induce more than one hard directions
 of magnetization parallel to the main plane whereas strength of the hard-axis anisotropy increases with size of the platelet
 along this axis. I propose to effectively describe the magnetostatic energy in the form of an integral over the stripe edge 
\begin{eqnarray}
E_{B}=\int_{-\infty}^{\infty}\left[-\frac{2}{\Lambda_{1}}(M^{2}-m_{x}^{2})
+\frac{2}{\Lambda_{2}}m_{z}^{2}\right]_{z=0}{\rm d}x
\label{E_B}
\end{eqnarray}
while e.g. the approach of Ref. \cite{che08} corresponds to $1/\Lambda_{1}=0$. By analogy
 to finite platelets, this coefficient is expected to scale with the stripe width following $\Lambda_{1}\propto\lambda^{2}/w$.
 More detailed estimation of $E_{B}$ is performed in Appendix A.

I mention that if $x$ direction was not one of the local hard axes at the boundaries of a wire (stripe) of finite length,
 the DW would be unstable due to a magnetostatic field at the wire (stripe) ends \cite{car10}.
 This local anisotropy prevents spontaneous DW motion towards one of the wire ends.
  
Inserting (\ref{profile1a})-(\ref{profile1b}) into the Hamiltonian (\ref{hamiltonian}), one arrives at 
\begin{eqnarray}
E_{0}(\theta,n)=2JMw
\sqrt{\frac{\beta_{1}}{J}[1+{\rm tan}^{2}(\theta)]+\frac{\pi^{2}n^{2}}{w^{2}}}.
\label{E_0}
\end{eqnarray} 
Evaluating $E_{B}$, I divide it into two parts ($E_{B}=E_{B1}+E_{B2}$); 
\begin{eqnarray}
E_{B1}(\theta,n)&\equiv&-\frac{2M^{2}}{\Lambda_{1}}\int_{-\infty}^{\infty}
{\rm sech}^{2}\left\{\frac{n\pi x}{w[1+\tan^{2}(\theta)]}\right\}{\rm d}x,
\nonumber\\
E_{B2}(\phi,\theta,n)&\equiv&\frac{2M^{2}}{\Lambda_{2}}\int_{-\infty}^{\infty}
{\rm sin}^{2}\left\{\phi-\frac{n\pi\tan(\theta)x}{w[1+\tan^{2}(\theta)]}\right\}
\nonumber\\
&&\times{\rm sech}^{2}\left\{\frac{n\pi x}{w[1+\tan^{2}(\theta)]}\right\}{\rm d}x.
\label{boundary_energy}
\end{eqnarray}
The minimization of $E_{B}(\phi,\theta,n)$ leads to $\phi=0,\pi$ independently of other parameters of the DW.  
 These values of $\phi$ correspond to the presence of halfvortices (halfantivortices) at the stripe edges (as shown in Fig. 1).
 In the regime of narrow stripes $w/\tau\sim 1$, $\Lambda_{2}\ll\Lambda_{1}$, ($1/\Lambda_{2}$ is small, thus,
 $E_{0}$ dominates over $E_{B}$), via minimization of $E_{0}(\theta,n)$ the smallest
 possible value of $n$ is preferable while the minimization of both $E_{0}(\theta,n)$, $E_{B}(0,\theta,n)$ points out 
 $\theta=0,\pi$ to be preferred. Increase of the stripe width $w$ with fixed thickness
 $\tau$ results in transition to the regime $\Lambda_{2}>\Lambda_{1}$. Furthermore, since $\beta_{1}\ll J/w^{2}$,
 from (\ref{E_0}), $E_{0}(\theta,n)\approx E_{0}(n)$ is independent
 of $w$, while, from (\ref{boundary_energy}), $E_{B}\propto w/\Lambda_{1}$. Hence, for big-enough $w$, $E_{B}$ becomes
 comparable to $E_{0}$. By the minimization of $E_{B1}(\theta,n)$, the biggest possible value of $\tan^{2}(\theta)$ and
 the smallest value of $n$ are preferable, whereas, for $\theta\neq 0,\pi$, the minimization of $E_{B2}(0,\theta,n)$ indicates
 big values of $n$ to be preferable. Hence, the transition from the DW state of $n=1$, $\theta=0,\pi$ to a state
 of $n=2$, $\theta\neq 0$ takes place with increase of $w$. For $n=2$, the condition $\tan^{2}(\theta)>1$ has to
 be satisfied in order that $E_{B}(0,\theta,2)<E_{B}(0,0,1)$.

The state of $n=1$ and $\theta=0,\pi$ corresponds to $k^{''}=q^{'}=0$, $|q^{''}|=\pi/w$, $|k^{'}|=\sqrt{\pi^2/w^2+\beta_{1}/J}$
 and it is called a {\it transverse DW}. With relevance to the state of $n=2$, I take $\beta_{1}=0$ for simplicity,
 and $|\theta|$ to be close to its infimum $|\theta|=\pi/4$. The resulting magnetization structure
 corresponds to $|q^{'}|=|q^{''}|=|k^{'}|=|k^{''}|=\pi/w$ and one calls it a {\it vortex DW}.
 The polarity of vortex (transverse) DW, (the magnetization orientation in the center of vortex/halfvortex, parallel
 or antiparallel one to $y$ axis), is determined by value of $\phi$ while $q^{''}/k^{'}=\pm 1$ determines its chirality,
 (the direction of magnetization rotation in the stripe plane in the vortex or antivortex cores, clockwise or anticlockwise one). 

Treatment of the ordering of soft-magnetic stripe within the present local approximation of micromagnetics provides
 a unified description of transverse and vortex DWs which is consistent with previously known exact DW solution 
 for the so-called exchange dominated regime of ultra-thin nanostripes ($n=1$) \cite{che05,che08}, and for 2D stripes
 with exchange interactions of Ref. \cite{met10}, ($n=2$).

\section{Dynamical transformations of domain wall}

The application of a longitudinal magnetic field (${\bf H}\neq 0$) enforces the DW translation along the wire (stripe)
 whose direction (parallel or antiparallel to $x$ axis) is determined by the decrease of Zeeman energy with time.
 Two regimes of the field intensity,
 separated by a critical (Walker breakdown) value $H_{W}$, have to be distinguished. For weak field $|H_{x}|<H_{W}$,
 the dynamics is restricted to the stationary DW translation, whereas, for $|H_{x}|>H_{W}$, the magnetization rotation 
 about the long axis of the stripe stripe is allowed. The phenomenon of Walker breakdown has been predicted by solving 
 the LLG equation in the framework of 1D model with two-axis anisotropy \cite{sch74}. Although, numerical estimations 
 of $H_{W}$ for nanostripes show invalidity of the 1D model to the systems under study, the field- and
 current-induced Walker breakdowns are observed in them \cite{bea07}. 

In order to include the breakdown phenomenon into the description of flat magnets, I discriminate between "weakly dissipative" and
 "purely relaxational" dynamical regimes. By analogy to the construction of a time-dependent Ginzburg-Landau equation
 in the theory of dynamic critical phenomena, in the "weakly dissipative" regime, the evolution is governed by Eq. (\ref{LLG})
 while the "purely relaxational" dynamics corresponds to a modified [by changing the LHS of (\ref{LLG}) into zero] equation
 of motion. I claim the conservative kinetic term (the LHS) of (\ref{LLG}) to be irrelevant to the case of  
 a weak-field (-current) since such a field induces a low-energy spin excitations that have to be overdamped 
 in an anisotropic medium \cite{hal72}. In the stripe model of Sec. II, the anisotropy is local (it is introduced
 via boundary conditions) and the mechanism of breakdown accompanied by disappearance of the kinetic term in (\ref{LLG})
 is similar to overdamping low-energy excitations (the central peak in excitation spectra) of isotropic systems doped
 with anisotropy centers \cite{hal76,aep84}.

Searching for the field-driven DW evolution, I modify the ansatz (\ref{solution_1}) into 
\begin{eqnarray}
f=1,\hspace*{2em}g=u{\rm e}^{kx+qz-lt}.
\label{ansatz_t}
\end{eqnarray} 
 With relevance to the "weakly dissipative" dynamical regime of $|H_{x}|>H_{W}$, via (\ref{secondary-eq}), for $k$ and $q$
 of the previous section, I find  
\begin{eqnarray}
l=\frac{-\gamma H_{x}}{{\rm i}+\alpha}.
\label{frequency_H}
\end{eqnarray} 
The imaginary part of $l$ is equal to the frequency of the magnetization rotation about $x$ axis. This rotation 
 will be shown to imply additional oscillations of the orientation of DW in the stripe plane due to effect
 of the magnetostatic field at the stripe boundary. Below the breakdown, for $|H_{x}|<H_{W}$, the DW solution
 to the "purely relaxational" secondary evolution equation [the LHSs of (\ref{secondary-eq}) are changed into zero]
 corresponds to $l=\gamma H_{x}/\alpha$. In both the cases of "weakly dissipative" and "purely relaxational" dynamics,
 motion of the cross-tie DWs can be thought of as a translation of vortices and antivortices of the DW texture.
 For $|H_{x}|<H_{W}$, the velocity of this motion is constant and oriented along the stripe, whereas, for $|H_{x}|>H_{W}$, 
 its direction is not conserved. In the last case, since directions of the vortex (antivortex) translation
 deviate from $x$ axis, value of the energy $E_{B}$ evolves, which results in an evolution of the DW parameters $\theta$, $n$.
 In order to determine $\theta(t)$, $n(t)$, I minimize
 the energy $E_{0}[\theta(t),n(t)]+E_{B}[\gamma H_{x}t/(1+\alpha^{2}),\theta(t),n(t)]$ with respect to these functions. 
 Since I assume $\theta(t)$, $n(t)$ to be independent of spatial variables, the DW stays in a cross-tie structure during
 whole the evolution time, which is due to domination of the exchange interactions over the dipolar ones. 
 Hence, the density of Bloch lines (the distance between vortices) and the DW width are uniform along the wall \cite{slo72},
 however, unlike in theories of the Bloch wall and Bloch line motion in platelets (which are valid to small propagation
 distances) \cite{slo73}, they are not conserved as far as $|H_{x}|>H_{W}$.

The constraint
\begin{eqnarray} 
n(t)/\{1+\tan^{2}[\theta(t)]\}=1
\label{constraint_3}
\end{eqnarray}
corresponds to transitions between states $(n=1,\theta=0$ or $\theta=\pi)$, $(n=2,|\theta|=\pi/4)$, $\ldots$, 
 transverse DWs and (multi-)vortex DWs. The motivation for writing (\ref{constraint_3}) is following. Along any straight
 $z=z_{0}\in(0,w)$, the magnetization component $m_{x}(x,z_{0},t)$ sustains a shift without changing its profile, 
 $\{m_{x}(x,z_{0},t)=m_{x}[x-x_{0}(t),z_{0},0]\}$. It is because the applied longitudinal field ${\bf H}$ drives
 only rotation of the remaining magnetization components about $x$ axis while any straight $z=z_{0}$ does not intersect
 the stripe edges where a local anisotropy is present. This condition is fulfilled when the DW parameter
 $k^{'}=\pm q^{''}$ is conserved and $q^{'}$ is transformed into $q^{'}(t)=\mp k^{''}(t)=k^{'}\tan[\theta(t)]$,
 which is equivalent to (\ref{constraint_3}). Inserting the unperturbed magnetization components     
\begin{eqnarray} 
m_{+}(x,z,t)&=&M{\rm e}^{{\rm i}(-l^{''}t+k^{''}x+q^{''}z)}{\rm sech}[k^{'}x+q^{'}z-l^{'}t],
\nonumber\\
m_{x}(x,z,t)&=&-M{\rm tanh}[k^{'}x+q^{'}z-l^{'}t],
\end{eqnarray}
(with $l^{'}\equiv{\rm Re}l$, $l^{''}\equiv{\rm Im}l$) into the sum $E_{0}+E_{B}$, for $\beta_{1}\approx 0$
 and $|l^{'}/l^{''}|=\alpha\ll 1$, and applying (\ref{constraint_3}), one finds 
\begin{eqnarray}
E_{0}(a)&=&2JM\pi(1+a^{2}),
\nonumber\\
E_{B}(a,t)&=&-\frac{2M^{2}}{\Lambda_{1}}\int_{-\infty}^{\infty}
{\rm sech}^{2}\left(\frac{\pi x}{w}\right){\rm d}x
\nonumber\\
&&+\frac{2M^{2}}{\Lambda_{2}}\int_{-\infty}^{\infty}
{\rm sin}^{2}\left\{\gamma H_{x}t/(1+\alpha^{2})-\frac{\pi ax}{w}\right\}
\nonumber\\
&&\times{\rm sech}^{2}\left(\frac{\pi x}{w}\right){\rm d}x,
\label{energy_a-t}
\end{eqnarray}
where, $a\equiv{\rm tan}(\theta)$. The variable $a$ is a genuine variational parameter for the present extremum problem,
 ($a=0$ corresponds to the transverse DWs, $a=\pm1$ to the vortex DWs). For some arbitrary chosen values
 of $J$, $\Lambda_{1}$, $\Lambda_{2}$, $w$, the energy function $E_{0}(a)+E_{B}(a,t)$ is plotted in Fig. 2a.
 The corresponding density plot of the derivative of $E_{0}(a)+E_{B}(a,t)$ over $a$ is presented in Fig. 2b where the contour
 $\partial[E_{0}(a)+E_{B}(a,t)]/\partial a=0$ is shown. This contour indicates three extremal curves $a(t)$ which relate
 to different trajectories of vortices and antivortices of the DW structure. The straight $a(t)=0$ does not minimize
 $E_{0}(a)+E_{B}(a,t)$, as seen from Fig. 2a, while the magnetization rotation about a constant direction with the frequency
 $|l^{''}|=\gamma|H_{x}|/(1+\alpha^{2})$ corresponds to $a(t)\ge 0$ or $a(t)\le 0$. The trajectories of vortices
 and antivortices in the DW texture are sketched in Fig. 3, where periodically created and annihilated transverse and vortex
 (antivortex) DWs are indicated with thin solid lines.

{\it It is seen from (\ref{energy_a-t}) that the dependent on $\Lambda_{1}$ contribution to the energy of the DW 
 [$E_{B1}$ of (\ref{boundary_energy})] does not play any role in the dynamical transformation of the texture since
 it is neither dependent on the variational parameter $a$ nor on time.}
 
\begin{figure}
\unitlength 1mm
\begin{center}
\begin{picture}(175,78)
\put(2,-5){\resizebox{65mm}{!}{\includegraphics{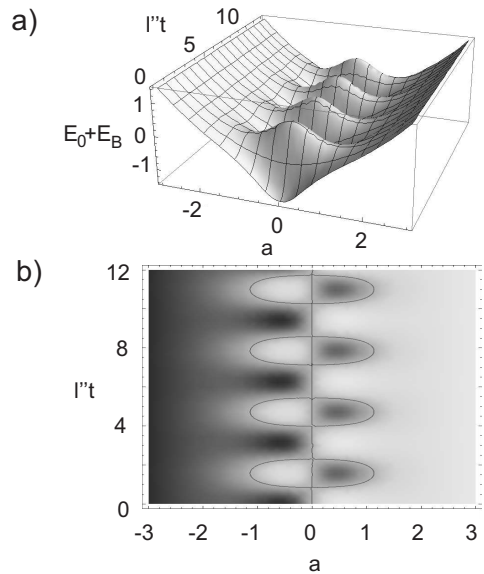}}}\end{picture}
\end{center}
\caption{a) Sum of the exchange and boundary energies $E_{0}+E_{B}$ with dependence on DW direction and time
for arbitrary chosen values of the parameters $J\Lambda_{1}/(Mw)=0,026$, $J\Lambda_{2}/(Mw)=0,020$. b) The 
corresponding density plot of the derivative of $E_{0}(a)+E_{B}(a,t)$ over $a$. The contour in the plot center represents
the solution to $\partial[E_{0}(a)+E_{B}(a,t)]/\partial a=0$.}
\end{figure}

\begin{figure}
\unitlength 1mm
\begin{center}
\begin{picture}(175,45)
\put(0,-5){\resizebox{85mm}{!}{\includegraphics{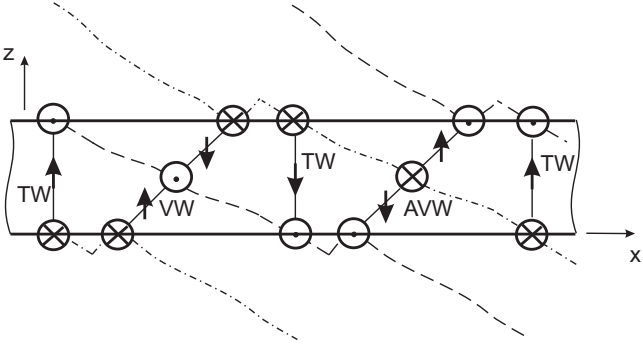}}}\end{picture}
\end{center}
\caption{Trajectories of vortices (dashed line) and antivortices (dash-dotted line) in the field-induced motion of a head-to-head DW.
The arrows indicate the magnetization directions in relevant DW areas.}
\end{figure}

\section{Vortex in circular dot and its transformations}

I study transformations of a magnetic vortex in a relatively-thick dot (of thickness $\tau>4l_{ex}$) in a pillar nanocontact 
 structure that contains a spin valve. The application of voltage to such a structure induces spin-polarized
 electric current through the dot plane and enforces cyclic magnetization motion. Unlike in very thin dots,
 where vortices are of small radii compared to dot radii because of strong effect of the magnetostatic field \cite{cap07},
 in thicker exchange-ordered dots, the radius of vortex core is comparable to the dot radius and the out-of-plane component
 of the magnetization at the dot boundary can be nonzero \cite{gus04}.

Since skyrmions are static solutions to the isotropic LLG equation in 2D \cite{bel75},
 they seem to be natural candidates to represent the vortices in dots. Indeed, in strongly flattened nanodots, ordering of vortex cores
 corresponds to skyrmion profiles (see Ref. \cite{gai10} for review of models of the centered vortex in the nanodot). Analytical approaches
 to the vortex-state dynamics of Refs. \cite{khv09,iva07} are based on transformations between local two-skyrmion or skyrmion-antiskyrmion
 textures embedded in an in-plane ordered curling background. They follow the observation that vortex in such a state can be shifted
 from the dot center and deformed due to interaction with another (virtual) vortex without production of surface magnetostatic
 charges \cite{met02}. However, in circular dots thick enough to allow using local approximation of the magnetostatics,
 the centered skyrmion is found not to be a stable state, (Appendix B), while any significant shift of the position 
 of static vortex from the dot center is not confirmed experimentally \cite{yu11}. Moreover, when the dot is built
 into a spin-valve-containing pillar structure, a weak magnetic anisotropy can be present whose easy axis is directed in the dot plane.
 In the presence of such an anisotropy, skyrmions are not solutions to the LLG equation while cross-tie DWs are. Therefore, I consider
 a single Bloch line of the cross-tie DW to be the center of a vortex in the magnetic dot, which is consistent with an observed
 breakdown of rotational symmetry of the dot ordering \cite{yu11,che09}. 
 
Studying magnetization configurations in the circular dots for ${\bf H}=0$, I use a static solution to (\ref{LLG}) in the form
 (\ref{profile1a})-(\ref{profile1b}) with $k^{''}=-q^{''}q^{'}/k^{'}=-q^{''}\tan(\theta)$,
 $k^{'2}-q^{''2}=\beta_{1}/\{J[1+\tan^{2}(\theta)]\}\approx 0$. Following the analysis of the previous section,
 the assumption of the orderings at the ends of any dot diagonal to be similar to each other leads to the relations
 $q^{''}=n\pi/\{2R[1+\tan^{2}(\theta)]\}=\pm k^{'}$, where $n=1,2,\ldots$, where $R$ denotes the dot radius.
 The exchange energy of such a dot state takes the form
\begin{eqnarray}
E_{0}(a)=JMk^{'2}\left(1+a^{2}\right)\int_{0}^{2\pi}\int_{0}^{R}
{\rm sech}^{2}\left\{k^{'}r
\right.\nonumber\\\Bigl.
\times[\sin(\varphi)+a\cos(\varphi)]\Bigr\}r{\rm d}r{\rm d}\varphi,
\end{eqnarray}
where $a\equiv\tan(\theta)$. Specific states characterized by $\phi=0$ or $\phi=\pi$ correspond to the presence
 of vortex or antivortex in the dot center. The boundary energy (of the magnetostatic origin)
\begin{eqnarray}
E_{B}=1/\Lambda_{2}\oint_{\partial S}({\bf m}\cdot{\bf n}^{'})^{2}{\rm d}l
\label{E_B_vortex}
\end{eqnarray}
(where ${\bf n}^{'}$ denotes the unitary vector
 outward to the dot boundary and $\Lambda_{2}\sim\lambda^{2}/[\tau|\log(\tau/2R)|]$ scales with the dot thickness $\tau$)
 is evaluated with 
\begin{eqnarray}
E_{B}(a,\phi)&=&\frac{M^{2}R}{\Lambda_{2}}\int_{0}^{2\pi}\left(-{\rm tanh}\left\{k^{'}R[\sin(\varphi)+a\cos(\varphi)]\right\}
\right.\nonumber\\
&&\times\sin(\varphi)
+{\rm sech}\left\{k^{'}R[\sin(\varphi)+a\cos(\varphi)]\right\}
\nonumber\\
&&\times
\sin\left\{\phi+k^{'}R[\cos(\varphi)-a\sin(\varphi)]\right\}\nonumber\\
&&\Bigl.
\times\cos(\varphi)\Bigr)^{2}{\rm d}\varphi.
\label{boundary_2}
\end{eqnarray}
The plot of energy $E_{0}(a)+E_{B}(a,0)$ corresponds to the cross section of 3D plot in Fig. 4a at $t=0$.
 Looking at this this cross section, one sees that absolute energy minimum corresponds to $|a|\approx 1$
 ($|\theta|\approx\pi/4$) which represents a vortex DW (Fig. 5a).
 
\begin{figure}
\unitlength 1mm
\begin{center}
\begin{picture}(175,80)
\put(0,-5){\resizebox{145mm}{!}{\includegraphics{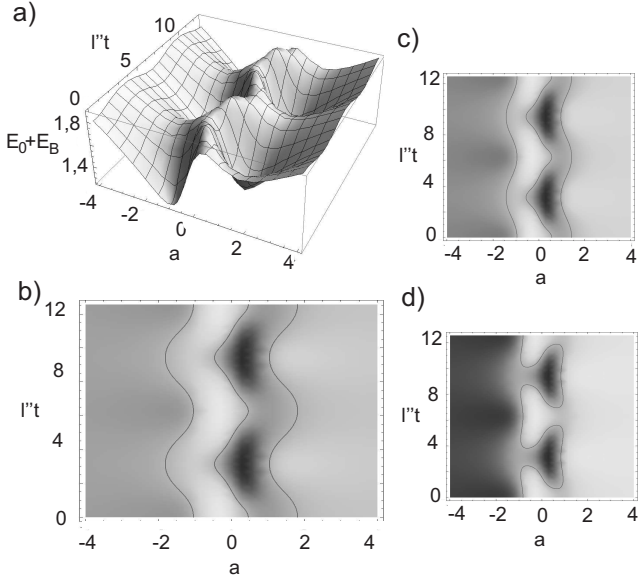}}}\end{picture}
\end{center}
\caption{a) Sum of the exchange and boundary energies $E_{0}+E_{B}$ with dependence on DW direction and time
for arbitrary chosen values of the parameters $J$, $R$, and $\Lambda_{2}=\Lambda$, [$J\Lambda/(MR)=0,022$].
b) The corresponding density plot of the derivative of $E_{0}(a)+E_{B}(a,t)$ over $a$. 
The contour in the plot center represents the solution to $\partial[E_{0}(a)+E_{B}(a,t)]/\partial a=0$.
c) The corresponding plot for $\Lambda_{2}=2\Lambda$, and d), for $\Lambda_{2}=4\Lambda$.}
\end{figure}

\begin{figure}
\unitlength 1mm
\begin{center}
\begin{picture}(175,108)
\put(0,-5){\resizebox{85mm}{!}{\includegraphics{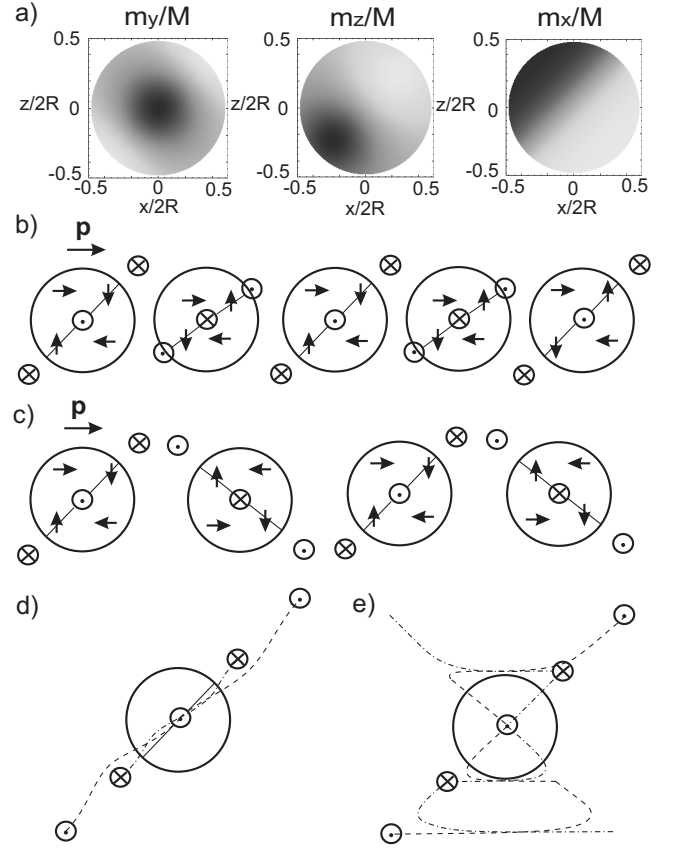}}}\end{picture}
\end{center}
\caption{In picture a), density plots of magnetization components of the dot state $|a|=1$, the grey scale is linear 
in the range [-1,1]. Below, schemes of consecutive textures (of the head-to-head DW type) and trajectories of vortices
(dashed lines) and antivortices (dash-dotted lines) during the transformation induced by a spin-polarized
current of the magnetic dot; b), d) $\Lambda_{2}<\Lambda_{c}$, c), e) $\Lambda_{2}>\Lambda_{c}$. 
The arrows in b) and c) indicate the magnetization alignment in relevant dot areas.}
\end{figure}

One includes an electric current perpendicular to the dot plane and spin-polarized in this plane via adding the spin-transfer torque 
\begin{eqnarray}
\frac{\sigma}{M}{\bf m}\times({\bf m}\times{\bf p})
\nonumber
\end{eqnarray}
to the RHS of (\ref{LLG}). Here $\sigma$ is proportional to the current intensity and ${\bf p}$ denotes its spin polarization
 ($|{\bf p}|=1$) \cite{khv09,iva07,slo96}. Applying the transformation (\ref{transform}), for $\beta_{1}\approx 0$ and ${\bf p}=(1,0,0)$, 
 one arrives at the secondary equations of motion
\begin{eqnarray}
-f{\rm i}D_{t}f^{*}\cdot g&=&f\left[\alpha D_{t}+J(D_{x}^{2}+D_{z}^{2})\right]f^{*}\cdot g 
\nonumber\\
&&+Jg^{*}(D_{x}^{2}+D_{z}^{2})g\cdot g-{\rm i}\sigma|f|^{2}g,
\nonumber\\
-g^{*}{\rm i}D_{t}f^{*}\cdot g&=&g^{*}\left[+\alpha D_{t}-J(D_{x}^{2}+D_{z}^{2})\right]f^{*}\cdot g
\nonumber\\
&&-Jf(D_{x}^{2}+D_{z}^{2})f^{*}\cdot f^{*}-{\rm i}\sigma|g|^{2}f^{*}.
\label{secondary-eq-2}
\end{eqnarray}
These equations can be obtained from (\ref{secondary-eq}) via change of $\gamma H_{x}$ into ${\rm i}\sigma$, thus,
 insertion of the ansatz (\ref{ansatz_t}) into (\ref{secondary-eq-2}) leads [by analogy to (\ref{frequency_H})] to 
\begin{eqnarray}
l=\frac{\sigma}{-1+{\rm i}\alpha}.
\label{frequency_sigma}
\end{eqnarray} 
In the weak current regime, below the Walker breakdown $\sigma<\sigma_{W}$, the LHSs of (\ref{secondary-eq-2})
 should be changed into zero and then
\begin{eqnarray}
l=\frac{-{\rm i}\sigma}{\alpha}={\rm i}l^{''}.
\end{eqnarray}
Since, in this regime, the real part of $l$ is equal to zero, unlike magnetic field, the polarized current does not
 induce any translation of the DW while it is responsible for rotation of the magnetization in the wall area about $x$ axis.

Including the magnetostatic field at the dot boundary, I analyze vortex-state dynamics as an evolution between different
 cross-tie DWs, thus, following the method of previous section, I look for the time dependence of the DW parameters
 $\theta$, $n$. I use the constraint (\ref{constraint_3}) which is motivated by considering the $m_{x}$ magnetization
 component along the straight $z=0$. For $\sigma<\sigma_{W}$, this component is conserved [$m_{x}(x,0,t)=m_{x}(x,0,0)$]
 upon application of a longitudinally-polarized current because this current drives the dynamics of the remaining magnetization
 components only, thus, it does not induce deviation of the magnetization from the local easy plane $yz$ at the ends
 of the dot diagonal $z=0$ [at $(x,z)=(\pm R,0)$]. On lines $z=z_{0}\neq 0$, the profile of $m_{x}$ is also conserved
 while its center sustains a shift $\{m_{x}(x,z_{0},t)=m_{x}[x-x_{0}(t),z_{0},0]\}$ because the current induces a deviation
 of ${\bf m}$ from an easy plane at $(x,z)=(\pm\sqrt{R^{2}-z_{0}^{2}},z_{0})$. With the constraint (\ref{constraint_3}),
 transforming the texture parameter $q^{'}$ into $q^{'}(t)=\mp k^{''}(t)=k^{'}\tan[\theta(t)]$, I substitute $\phi$ 
 in (\ref{boundary_2}) with $l^{''}t$ and plot the energy $E_{0}(a)+E_{B}(a,t)$ for some arbitrary values of $J$, $\Lambda_{2}$,
 $R$ (Fig. 4a) as well as its derivative over $a$ (Figs. 4b-4d). The contour $\partial[E_{0}(a)+E_{B}(a,t)]/\partial a=0$ indicates
 the trajectory $a(t)={\rm tan}[\theta(t)]$. There exists a critical value $\Lambda_{c}$ such that the function $a(t)$
 is discountinuous for $\Lambda_{2}>\Lambda_{c}$. Jumps of $a(t)$ are connected to changes of its sign as follows from Fig. 4d.
 For $\Lambda_{2}<\Lambda_{c}$, sign of $a(t)$ is constant (see Figs. 4b,4c). A number of consecutive textures created during
 a single period of the vortex-structure transformation is visualized in Figs. 5b, 5d together with shape of the corresponding
 vortex and antivortex trajectories (Figs. 5c,5e).

\begin{figure}
\unitlength 1mm
\begin{center}
\begin{picture}(175,18)
\put(0,-5){\resizebox{85mm}{!}{\includegraphics{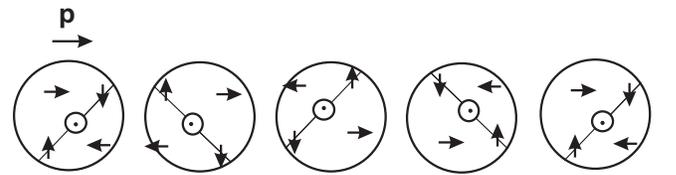}}}\end{picture}
\end{center}
\caption{Consecutive textures (of the head-to-head DW type) in a strongly driven ($\sigma>\sigma_{W}$) circular dot.
The arrows indicate the magnetization directions in relevant dot areas.}
\end{figure}

For a current of high intensity $\sigma>\sigma_{W}$, the relation (\ref{frequency_sigma}) leads to prediction of a texture 
 motion along the polarization direction ${\bf p}=(1,0,0)$ with the velocity $v={\rm Re}l/k^{'}=\sigma/[k^{'}(1+\alpha^{2})]$.
 In the confined geometry, however, this motion is suppressed when a decrease of energy due to the spin-transfer-induced
 shift of the vortex core from the dot center equals an increase of the boundary energy $E_{B}$. A non-zero value
 of $l^{''}$ corresponds to the magnetization rotation in the DW area (in the vortex core) with frequency 
 $l^{''}=\sigma\alpha/(1+\alpha^{2})$ about $x$ axis. In order that such a rotation would not induce any instability 
 of the structure (the boundary and exchange energies were constant), it should be accompanied by a precession 
 of the vortex (the DW texture) in plane of the dot shown in Fig. 6. The above determined frequency $l^{''}$
 of the vortex precession in a circular dot coincides with the relevant formula obtained in Ref. \cite{cho09}
 by solving the Thiele equation where it has been verified with micromagnetic simulations.
 The radius of the vortex precession can be small compared to the dot diameter \cite{yu11,che09}.

\section{Conclusions}

Cyclic transformations between transverse and vortex DWs in a strong longitudinal magnetic field have been described 
 with relevance to magnetic stripes with dominance of the exchange interactions over magnetostatic ones.  
 Similar effect has been observed in experiments and simulations for very thin magnetic stripes (of the thickness 
 of up to four magnetostatic exchange lengths, which corresponds to 20nm for permalloys), where magnetostatic
 field is strong in bulk and dynamical distortions of the DWs take place. Such DW distortions cannot
 be included into the present approach. The predicted dynamics can be verified with thicker
 stripes whose DW structures are expected to be more robust.

A specific feature of the DW dynamics above the Walker breakdown are consecutive pinnings and depinnings of a single halfvortex
 or an halfantivortices to the alternating stripe edges during the transformations of transverse DW into vortex DW and
 vice versa \cite{uhl09}. It is similar to the behavior shown in Fig. 3, however, since there is no any DW distortion
 in the present study, the half(anti)vortices are not strongly pinned to one of the stripe edges while they move away
 from the stripe on a small distance. Another difference with respect to DW dynamics in very thin nanostripes is lack
 of a simulated and explained with Thiele equation effect of dynamical reversal of the vortex-core polarity during
 its collision with the stripe edge \cite{lee07,cla08,min10}. It is because of different structures
 of the "magnetostatically-ordered" DWs of the cited works compared to the ones of "exchange-ordered" walls.
 In thin nanostripes, the field-excited texture can take the form
 of a transverse DW of total skyrmion number $G=0$, ($G\equiv\sum_{a}p_{a}q_{a}$ with $p_{a}=\pm 1$ denoting the polarity
 and $q_{a}=\pm 1/2,\pm 1$ denoting the chirality of $a$th defect in the DW texture \cite{tre07}), or a vortex DW of $G=\pm 1$.
 It means that the polarities of halfvortex and halfantivortex in the transverse wall could be the same while the polarities 
 of halfantivortices at the edges of a vortex DW can be opposite. In the present considerations, only values of $G=\pm 1$
 for the transverse DWs and $G=\pm 2$ for the vortex DWs are allowed, (the DWs are of a rigid cross-tie structure).
  
Two intensity regimes of the spin-polarized electric current are found to relate to periodic transformations of the vortex states
 in circular dots with frequencies of different orders of magnitude. Since $\alpha\ll 1$, the increase of the oscillation
 frequency $l^{''}$ with $\sigma$ is much faster for currents weaker than a threshold value $\sigma<\sigma_{W}$,
 (when $l^{''}=\sigma/\alpha$), than for stronger currents $\sigma>\sigma_{W}$, [when $l^{''}=\sigma\alpha/(1+\alpha^{2})$].
 The regime $\sigma>\sigma_{W}$ corresponds to a vortex gyration in the dot plane, (this regime of the current intensities
 is used to be studied with relevance to the dc-induced generation of microwaves), while the regime $\sigma<\sigma_{W}$
 corresponds to a vortex motion along an open trajectory. Most frequently, the vortex motion has been studied dealing with
 very thin dots with current above the threshold intensity. Schematic snapshots of the gyrating vortex of Fig. 6 can
 be compared to the ones observed in relatively thick (30nm) and of big diameter (3$\mu$m) dots during the magnetic-field-induced
 gyration [since descriptions of both the processes are similar except change of the parameter (\ref{frequency_sigma})
 into (\ref{frequency_H})] \cite{che09}. Small number of data on the vortex dynamics below the threshold
 is available. Similar (non-gyrotropic) evolution of the vortex state to the one of Fig. 5a has been observed in hexagonal dots 
 (of 50nm thickness) upon application of a nanosecond electromagnetic pulse (such a pulse induces a spin-polarized current
 and a longitudinal magnetic field simultaneously) \cite{shi10}. 
 
\section*{Acknowledgements}

This work was partially supported by Polish Government Research Founds for 2010-2012 in the framework of Grant No. N N202 198039.

\appendix
\section{Estimation of boundary energy}

The magnetostatic energy of a magnetic element contains contributions that relate to interactions of surface charges, 
 volume charges, and interaction between surface and volume ones 
\begin{eqnarray}
E_{MS}&=&\int\int\frac{\rho({\bf x})\rho({\bf x}^{'})}{|{\bf x}-{\bf x}^{'}|}{\rm d}V({\bf x}){\rm d}V({\bf x}^{'})
\nonumber\\
&&+\int\int\frac{\sigma({\bf x})\sigma({\bf x}^{'})}{|{\bf x}-{\bf x}^{'}|}{\rm d}S({\bf x}){\rm d}S({\bf x}^{'})
\nonumber\\
&&+\int\int\frac{\sigma({\bf x})\rho({\bf x}^{'})}{|{\bf x}-{\bf x}^{'}|}{\rm d}S({\bf x}){\rm d}V({\bf x}^{'}),
\end{eqnarray}
where $\rho=-\nabla\cdot{\bf m}$, $\sigma={\bf n}\cdot{\bf m}$. Influence of magnetostatic interactions on the 
 magnetization of the main body of any ferromagnetic platelet decreases with increase of its thickness, while 
 it is not so in terms of magnetization of the platelet boundary. Following Ref. \cite{bro65}, reducing one of
 the spatial dimensions with relevance to flat systems of thickness $\tau$ and neglecting volume and base-surface
 terms, the above expression is transformed into the energy of the boundary of 2D system (up to the multiplayer $\tau$)
\begin{eqnarray}
\tau E_{B}&=&\tau^{2}\int_{\partial S}\int_{\partial S}
\sigma({\bf x})\sigma({\bf x}^{'}){\rm ln}(|{\bf x}-{\bf x}^{'}|/\tau){\rm d}l({\bf x}){\rm d}l({\bf x}^{'})
\nonumber\\
&&+\tau^{2}\int_{\partial S}\int_{S_{base}}
\left[\sigma({\bf x})\rho({\bf x}^{'})+\rho({\bf x})\sigma({\bf x}^{'})\right]
\nonumber\\
&&\times
{\rm ln}(|{\bf x}-{\bf x}^{'}|/\tau)
{\rm d}l({\bf x}){\rm d}S({\bf x}^{'}).
\end{eqnarray}
Here $S_{base}$ denotes the surface of the platelet base.

For a circular dot, $\rho({\bf x})$ is not defined at its boundary, thus, I neglect the second term and obtain
\begin{eqnarray}
\tau E_{B}=\tau^{2}\int_{\partial S}\int_{\partial S}({\bf n}\cdot{\bf m})({\bf x})({\bf n}\cdot{\bf m})({\bf x}^{'})
\nonumber\\
{\rm ln}(|{\bf x}-{\bf x}^{'}|/\tau){\rm d}l({\bf x}){\rm d}l({\bf x}^{'})
\nonumber\\
\sim
2\pi R\tau^{2}{\rm ln}(R/\tau)\int_{\partial S}({\bf n}\cdot{\bf m})^{2}({\bf x}){\rm d}l({\bf x}).
\end{eqnarray}

For any DW in the stripe, $\rho({\bf x})\sim-\partial m_{x}/\partial x\sim\left[M^{2}-m_{x}^{2}\right]/(M\delta)$
 with a constant $\delta$ close to DW width, \{since
 $-\partial m_{x}/\partial x=M\partial[{\rm tanh}(x/\delta+cz/\delta)]/\partial x=[M^{2}-M^{2}{\rm tanh}^{2}(x/\delta+cz/\delta)]/(M\delta)$, 
 and $|\partial m_{z}/\partial z|\le|\partial m_{x}/\partial x|$\}. I estimate the boundary energy with
\begin{eqnarray}
\tau E_{B}&=&\tau^{2}\int_{\partial S}\int_{\partial S}
({\bf n}\cdot{\bf m})({\bf x})({\bf n}\cdot{\bf m})({\bf x}^{'}){\rm ln}(|{\bf x}-{\bf x}^{'}|/\tau)
\nonumber\\
&&\times
{\rm d}l({\bf x}){\rm d}l({\bf x}^{'})
-\tau^{2}\int_{\partial S}\int_{S_{base}}
\left[({\bf n}\cdot{\bf m})({\bf x})\frac{\partial m_{x}}{\partial x}({\bf x}^{'})
\right.
\nonumber\\
&&\left.
+\frac{\partial m_{x}}{\partial x}({\bf x})({\bf n}\cdot{\bf m})({\bf x}^{'})\right]
{\rm ln}(|{\bf x}-{\bf x}^{'}|/\tau)
{\rm d}l({\bf x}){\rm d}S({\bf x}^{'})
\nonumber\\
&\sim&
2\delta\tau^{2}{\rm ln}(\delta/\tau)\int_{-\infty}^{\infty}m_{z}^{2}(x,0,0){\rm d}x
\nonumber\\
&&-2\kappa w\tau^{2}{\rm ln}(\delta/\tau)\int_{-\infty}^{\infty}\left[M^{2}-m_{x}^{2}(x,0,0)\right]{\rm d}x,
\label{DW_magnetostatics}
\end{eqnarray}
where $\kappa w$ corresponds to an effective thickness of the surface layer of the stripe edge over which the magnetization 
 is independent of normal coordinate $z$, ($\kappa\ll 1$ and $\kappa\propto\delta$). Since $\delta\sim w$, one arrives at $E_{B}$ of (\ref{E_B}).

In order that boundary conditions to the LLG equation (within 2D approximation) were determined, one requires the relation
 $l_{ex}^{2}/w^{2}\ll\tau/w{\rm ln}(w/\tau)$ to be satisfied following Kurzke \cite{kur05}, while, unlike it is 
 assumed in his thin-film-limit approach, another condition $\tau/w\ll l_{ex}^{2}/w^{2}$ is not fulfilled in present considerations.
 In Ref. \cite{koh05}, the last inequality was used in order to show that the bulk-bulk and bulk-boundary terms 
 of the magnetostatic energy vanish faster than the boundary-boundary term in the thin-film limit, thus, they are negligible
 for very thin structures, (also, then, a strong easy-plane anisotropy justifies using XY model). 
 DWs of this "exchange" limit are described with exact formula in Refs. \cite{che05,che08}.
 When $\tau>l_{ex}$ (the platelet is thick-enough), the bulk-boundary term can be non-negligible.
 Its importance grows with increase of lateral diameter of the platelet [stripe width, increase of the coefficient $\kappa$
 of (\ref{DW_magnetostatics})] with constant $\tau$. In most of experiments (and in the present paper), sizes of the system
 lie far from the Kurzke regime, approaching to the region $\tau/w\gg l_{ex}^{2}/w^{2}$. 

\section{Instability of skyrmion in circular dot}

I determine the energy of a skyrmion centered with respect to a circular magnetic dot using 
 a solution to (\ref{LLG}) for $\beta_{1}=0$, ${\bf H}_{x}=0$ (in absence of the anisotropy and
 external field), with the boundary conditions; $m_{y}\to-M$, $m_{z}+{\rm i}m_{x}\to 0$ with $\sqrt{x^{2}+z^{2}}\to\infty$. 
 Applying the Hirota method, via transforming magnetization to secondary dynamical variables 
\begin{eqnarray}
m_{z}+{\rm i}m_{x}=2M\frac{gf}{|f|^{2}+|g|^{2}},
\hspace*{0.5em}
m_{y}=M\frac{|f|^{2}-|g|^{2}}{|f|^{2}+|g|^{2}},
\end{eqnarray}
the skyrmion solution to the resulting equation (\ref{secondary-eq}) is found to be of the form 
\begin{eqnarray}
f=1,\hspace*{2em}g={\rm e}^{{\rm i}\eta+{\rm i}q\cdot{\rm arctan}(x/z)}\frac{\sqrt{x^{2}+z^{2}}}{{\cal R}},
\end{eqnarray}
where ${\cal R}$ is a characteristic texture width, $q=1$ (skyrmion) or $q=-1$ (antiskyrmion). 
 The exchange and magnetostatic energies of the skyrmion and antiskyrmion are evaluated with 
 (\ref{hamiltonian}) and (\ref{E_B_vortex})
\begin{eqnarray}
E_{0}({\cal R})&=&\frac{JM}{2}\int_{0}^{2\pi}\int_{0}^{R}
\frac{2/{\cal R}^{2}+16r^{2}/{\cal R}^{4}+2r^{4}/{\cal R}^{6}}{(1+r^{2}/{\cal R}^{2})^{4}}r{\rm d}r{\rm d}\varphi
\nonumber\\
&=&JM\pi\frac{R^{2}/{\cal R}^{2}+5R^{4}/{\cal R}^{4}+2R^{6}/{\cal R}^{6}}{(1+R^{2}/{\cal R}^{2})^{3}},
\end{eqnarray}
\begin{eqnarray}
E_{B}({\cal R},\eta)&=&\frac{M^{2}}{\Lambda_{2}}\int_{0}^{2\pi}
\frac{R^{2}/{\cal R}^{2}\cos^{2}[(1-q)\varphi-\eta]}{(1+R^{2}/{\cal R}^{2})^{2}}R{\rm d}\varphi.
\nonumber\\
&=&\left\{\begin{array}{lc}
\frac{M^{2}}{\Lambda_{2}}\frac{R^{3}/{\cal R}^{2}}{(1+R^{2}/{\cal R}^{2})^{2}}2\pi\cos^{2}(\eta) & q=1,\\
\frac{M^{2}}{\Lambda_{2}}\frac{R^{3}/{\cal R}^{2}}{(1+R^{2}/{\cal R}^{2})^{2}}\frac{\pi}{2} & q=-1.
\end{array}
\right.
\end{eqnarray}
Via minimization of $E_{0}({\cal R})+E_{B}({\cal R},\eta)$, infinitely big texture radius ${\cal R}$ is preferred, which leads to
 the instability of the centered skyrmion (antiskyrmion) state. 
 
It should be emphasized that the above conclusion is valid as far as the local approach is suitable. 
 Full evaluation of the magnetostatic energy can indicate stability of a centered vortex embedded
 in a curling in-plane magnetization state, especially for dots of high radius-to-thickness
 ratio \cite{gai10}. Such stable centered vortices are often called skyrmions because of skyrmion-like core magnetization,
 however, in a nomenclature widely used in different branches of physics, they are not skyrmions since they are not defects
 in the ferromagnetically ordered background.

\end{document}